\documentclass[aps,secnumarabic,nobalancelastpage,
nofootinbib,onecolumn,12pt]{revtex4}%
\usepackage{amsfonts}
\usepackage{amsmath}
\usepackage{graphics}
\usepackage{graphicx}
\usepackage{longtable}
\usepackage{url}
\usepackage{bm}
\usepackage{amssymb}%
\begin{document}
\title{Poisson-Fermi Formulation of Nonlocal Electrostatics in Electrolyte Solutions}
\author{Jinn-Liang Liu}
\affiliation{Institute of Computational and Modeling Science, National Tsing Hua
University, Hsinchu 300, Taiwan. E-mail: jlliu@mx.nthu.edu.tw}
\author{Dexuan Xie}
\affiliation{Department of Mathematical Sciences, University of Wisconsin-Milwaukee,
Milwaukee, WI 53201-0413, USA. E-mail: dxie@uwm.edu}
\author{Bob Eisenberg}
\affiliation{Department of Molecular Biophysics and Physiology, Rush University, Chicago,
IL 60612 USA. E-mail: beisenbe@rush.edu}
\date{\today }

\begin{abstract}

\end{abstract}
\maketitle

We present a nonlocal electrostatic formulation of nonuniform ions and water
molecules with interstitial voids that uses a Fermi-like distribution to
account for steric and correlation effects in electrolyte solutions. The
formulation is based on the volume exclusion of hard spheres leading to a
steric potential and Maxwell's displacement field with Yukawa-type
interactions resulting in a nonlocal electric potential. The classical
Poisson-Boltzmann model fails to describe steric and correlation effects
important in a variety of chemical and biological systems, especially in high
field or large concentration conditions found in and near binding sites, ion
channels, and electrodes. Steric effects and correlations are apparent when we
compare nonlocal Poisson-Fermi results to Poisson-Boltzmann calculations in
electric double layer and to experimental measurements on the selectivity of
potassium channels for K$^{+}$ over Na$^{+}$. The present theory links atomic
scale descriptions of the crystallized KcsA channel with macroscopic bulk
conditions. Atomic structures and macroscopic conditions determine complex
functions of great importance in biology, nanotechnology, and electrochemistry.

Continuum electrostatic theory is a fundamental tool for studying physical and
chemical properties of electrolyte solutions in a wide range of applications
in electrochemistry, biophysics, colloid science, and nanofluidics
\cite{LM03,H01,BW01,HB04,AA07,BS11}. For over a century, a great deal of
effort has been devoted to improving the Poisson-Boltzmann (PB) theory of
continuum models for a proper description of steric (or finite size) and
correlation (or nonlocal screening, polarization) effects in electrolytes
\cite{BK09,A96E13}. We present a continuum model with Fermi-like distributions
and global electrostatic screening of nonuniform ions and water molecules to
describe the steric and correlation effects, respectively, in aqueous
electrolyte solutions.

For an electrolytic system with $K$ species of ions, the entropy model
proposed in \cite{LE14} treats all ions and water of any diameter as
nonuniform hard spheres and regards the water as the $(K+1)^{\text{th}}$
species. It then includes the voids between these hard spheres as the
$(K+2)^{\text{th}}$ species so that the total volume $V$ of the system can be
calculated exactly by the identity%
\begin{equation}
V=\sum_{i=1}^{K+1}v_{i}N_{i}+V_{K+2}, \tag{1}%
\end{equation}
where $V_{K+2}$ denotes the total volume of all the voids, $v_{i}=4\pi
a_{i}^{3}/3$ that gives the volume of each sphere with radius $a_{i}$, and
$N_{i}$ is the total number of the $i^{\text{th}}$ species particles. In the
bulk solution, we have the bulk concentrations $C_{i}^{\text{B}}=\frac{N_{i}%
}{V}$ and the bulk volume fraction of voids $\Gamma^{\text{B}}=\frac{V_{K+2}%
}{V}$. Dividing the volume identity (1) by $V$, $\Gamma^{\text{B}}%
=1-\sum_{i=1}^{K+1}v_{i}C_{i}^{\text{B}}$ is expressed in terms of nonuniform
$v_{i}$ and $C_{i}^{\text{B}}$ for all particle species. If the system is
spatially inhomogeneous with variable electric or steric fields, as in most
biological and technological systems, the bulk concentrations then change to
concentration functions $C_{i}(\mathbf{r})$ that vary with positions, and
differ from their constant values $C_{i}^{\text{B}}$ at location $\mathbf{r}$
in the solvent domain $\Omega_{s}$. Consequently, the void volume fraction
becomes a function $\Gamma(\mathbf{r)}=1-\sum_{i=1}^{K+1}v_{i}C_{i}%
(\mathbf{r})$ as well.

For an electrolyte in contact with electrodes or containing a charged protein,
an electric field $\mathbf{E}(\mathbf{r})$ in the solvent domain $\Omega_{s}$
is generated by the electrodes, ionic (free) charges with a displacement field
$\mathbf{D}(\mathbf{r})$, and bound charges of polar water with a polarization
field $\mathbf{P}(\mathbf{r})$. In Maxwell's theory, these fields form a
constitutive relation%
\begin{equation}
\mathbf{D}(\mathbf{r})=\epsilon_{0}\mathbf{E}(\mathbf{r})+\mathbf{P}%
(\mathbf{r}) \tag{2}%
\end{equation}
that yields the Maxwell's equation $\nabla\cdot\mathbf{D}(\mathbf{r}%
)=\rho(\mathbf{r})=\sum_{i=1}^{K}q_{i}C_{i}(\mathbf{r})$, $\forall
\mathbf{r}\in\Omega_{s}$, where $\epsilon_{0}$ is the vacuum permittivity and
$q_{i}$ is the charge on each $i$ species ion \cite{Z13}. The electric field
$\mathbf{E}(\mathbf{r})$ is thus screened by water (in what might be called
Bjerrum screening) and ions (in Debye screening) in a correlated manner that
is usually characterized by a correlation (screening) length $\lambda$
\cite{HB04,S06,BS11}. The screened force between two charges in ionic
solutions (at $\mathbf{r}$ and $\mathbf{r}^{\prime}$ in $\Omega_{s}$) has been
studied extensively in classical field theory and is often described by the
screening kernel $G(\mathbf{r}-\mathbf{r}^{\prime})=\frac{e^{-\left\vert
\mathbf{r}-\mathbf{r}^{\prime}\right\vert /\lambda}}{4\pi\left\vert
\mathbf{r}-\mathbf{r}^{\prime}\right\vert }$ \cite{LM03}, which is called
Yukawa-type kernel in \cite{HB04,XJ12}, and satisfies the partial differential
equation (PDE) \cite{XJ12}%
\begin{equation}
-\Delta G(\mathbf{r}-\mathbf{r}^{\prime})+\frac{1}{\lambda^{2}}G(\mathbf{r}%
-\mathbf{r}^{\prime})=\delta(\mathbf{r}-\mathbf{r}^{\prime})\text{,
\ }\mathbf{r}\text{, }\mathbf{r}^{\prime}\in R^{3} \tag{3}%
\end{equation}
in the whole space $R^{3}$, where $\Delta=\nabla\cdot\nabla$ is the Laplace
operator with respect to $\mathbf{r}$ and $\delta(\mathbf{r}-\mathbf{r}%
^{\prime})$ is the Dirac delta function at $\mathbf{r}^{\prime}$. The
potential $\widetilde{\phi}(\mathbf{r})$ defined in $\mathbf{D}(\mathbf{r}%
)=-\epsilon_{s}\epsilon_{0}\nabla\widetilde{\phi}(\mathbf{r})$ thus describes
a local potential of free ions according to the Poisson equation%
\begin{equation}
-\epsilon_{s}\epsilon_{0}\nabla\cdot\nabla\widetilde{\phi}(\mathbf{r}%
)=\rho(\mathbf{r})\text{, \ }\forall\mathbf{r}\in\Omega_{s}\text{,} \tag{4}%
\end{equation}
where $\epsilon_{s}$ is a dielectric constant in the solvent domain. We
introduce a global electric potential $\phi(\mathbf{r})$ of the screened
electric field $\mathbf{E}(\mathbf{r})$ as a convolution of the local
potential $\widetilde{\phi}(\mathbf{r}^{\prime})$ with the global screening
kernel $G(\mathbf{r}-\mathbf{r}^{\prime})$ in the expression
\begin{equation}
\phi(\mathbf{r})=\int_{R^{3}}\frac{1}{\lambda^{2}}G(\mathbf{r}-\mathbf{r}%
^{\prime})\widetilde{\phi}(\mathbf{r}^{\prime})d\mathbf{r}^{\prime}\text{.}
\tag{5}%
\end{equation}
However, it would be too expensive to calculate $\phi(\mathbf{r})$ using this
equation. Multiplying Eq. (3) by $\widetilde{\phi}(\mathbf{r}^{\prime})$ and
then integrating over $R^{3}$ with respect to $\mathbf{r}^{\prime}$
\cite{XJ12}, we obtain
\begin{equation}
-\lambda^{2}\Delta\phi(\mathbf{r})+\phi(\mathbf{r})=\widetilde{\phi
}(\mathbf{r})\text{, \ }\mathbf{r}\in\Omega_{s}, \tag{6}%
\end{equation}
a PDE that approximates Eq. (5) in a sufficiently large domain $\Omega_{s}$
with boundary conditions $\phi(\mathbf{r})=\widetilde{\phi}(\mathbf{r})=0$ on
$\partial\Omega_{s}$ and describes the relation between global $\phi
(\mathbf{r})$ and local $\widetilde{\phi}(\mathbf{r})$ electric potentials.
From Eqs. (4) and (6), we obtain the fourth-order PDE%
\begin{equation}
\epsilon_{s}\epsilon_{0}\lambda^{2}\Delta(\Delta\phi(\mathbf{r}))-\epsilon
_{s}\epsilon_{0}\Delta\phi(\mathbf{r})=\rho(\mathbf{r}),\text{\ }\mathbf{r}%
\in\Omega_{s}\text{,} \tag{7}%
\end{equation}
that accounts for electrostatic, correlation, polarization, nonlocal, and
excluded volume effects in electrolytes with only one parameter $\lambda$.
Thus, when we set $\mathbf{E}(\mathbf{r})=-\nabla\phi(\mathbf{r})$, we can use
Eq. (2) to find the polarization field $\mathbf{P}(\mathbf{r})=\epsilon
_{s}\epsilon_{0}\lambda^{2}\nabla(\Delta\phi(\mathbf{r}))-(\epsilon
_{s}-1)\epsilon_{0}\nabla\phi(\mathbf{r})$. If $\lambda=0$, we recover the
standard Poisson equation (4) and the classical field relation $\mathbf{P}%
=\epsilon_{0}(\epsilon_{s}-1)\mathbf{E}$ with the electric susceptibility
$\epsilon_{s}-1$ (and thus the dielectric constant $\epsilon_{s}$) if water is
treated as an isotropic and linear dielectric medium \cite{Z13}.

We introduce a Gibbs free energy functional for the system as
\begin{align}
F(\mathbf{C},\phi)  &  =F_{el}(\mathbf{C},\phi)+F_{en}(\mathbf{C}%
)\text{,}\tag{8}\\
F_{el}(\mathbf{C},\phi)  &  =\frac{1}{2}\int_{\Omega_{s}}\rho\phi
d\mathbf{r}=\frac{1}{2}\int_{\Omega_{s}}\rho L^{-1}\rho d\mathbf{r}\text{,
}\nonumber\\
F_{en}(\mathbf{C})  &  =k_{B}T\int_{\Omega_{s}}\left\{  \sum_{i=1}^{K+1}%
C_{i}(\mathbf{r})\left(  \ln\frac{C_{i}(\mathbf{r})}{C_{i}^{\text{B}}%
}-1\right)  +\frac{\Gamma(\mathbf{r})}{v_{0}}\left(  \ln\frac{\Gamma
(\mathbf{r})}{\Gamma^{\text{B}}}-1\right)  \right\}  d\mathbf{r}%
\text{,}\nonumber
\end{align}
where $F_{el}(\mathbf{C},\phi)$ is an electrostatic functional, $F_{en}%
(\mathbf{C})$ is an entropy functional, $\mathbf{C}=\left(  C_{1}%
(\mathbf{r})\text{, }C_{2}(\mathbf{r})\text{,}\cdots\text{, }C_{K+1}%
(\mathbf{r}\mathbb{)}\right)  $, $v_{0}=\left(  \sum_{i=1}^{K+1}v_{i}\right)
/(K+1)$ an average volume, and $L^{-1}$ is the inverse of the self-adjoint
positive linear operator $L=\epsilon_{s}\lambda^{2}\Delta\Delta-\epsilon
_{s}\Delta$ \cite{XJ12}. Taking the variations of $F(\mathbf{C},\phi)$ at
$\phi$ gives Eq. (7). Taking the variations of $F(\mathbf{C},\phi)$ at
$C_{i}(\mathbf{r}) $ \cite{XJ12} yields Fermi-like distributions%
\begin{equation}
C_{i}(\mathbf{r})=C_{i}^{\text{B}}\exp\left(  -\beta_{i}\phi(\mathbf{r}%
)+\frac{v_{i}}{v_{0}}S^{\text{trc}}(\mathbf{r})\right)  \text{, \ \ }%
S^{\text{trc}}(\mathbf{r})=\ln\left(  \frac{\Gamma(\mathbf{r)}}{\Gamma
^{\text{B}}}\right)  , \tag{9}%
\end{equation}
for all $i=1,$ $\cdots,$ $K+1$ (ions and water), where $\beta_{i}=q_{i}%
/k_{B}T$ with $q_{K+1}=0$, $k_{B}$ is the Boltzmann constant, and $T$ is an
absolute temperature. The distribution is of Fermi type since it saturates.
All concentration functions $C_{i}(\mathbf{r})<\frac{1}{v_{i}}$ \cite{LE14},
i.e., $C_{i}(\mathbf{r})$ cannot exceed the maximum value $1/v_{i}$ for any
arbitrary (or even infinite) potential $\phi(\mathbf{r})$ in the domain
$\Omega_{s}$. In these Fermi distributions, it is impossible for a particle
volume $v_{i}$ to be completely filled with particles, i.e., it is impossible
to have $v_{i}C_{i}(\mathbf{r})=1$ (and thus $\Gamma(\mathbf{r)}=0$) since
that would yield $S^{\text{trc}}(\mathbf{r})=-\infty$ and hence $v_{i}%
C_{i}(\mathbf{r})=0$, a contradiction. For this reason,\textit{\ we must
include the void as a separate species if water and ions are all treated as
hard spheres} \cite{LE14}. Here we do represent water and ions as spheres. Our
approach allows other shapes to be used as well.

The nonlocal Poisson-Fermi (PF) Eqs. (7) and (9) have new features, of some importance.

(i) The Fermi-like distribution of uniform spherical ions with voids in ionic
liquids was first derived by Bazant et al. \cite{BS11,BK09} using Bikerman's
lattice model \cite{B42}. The entropy functional in \cite{BS11} involves a
reciprocal of a minimum volume $v$ with a volume fraction $\Phi$ that is an
empirical fitting parameter and may have to be unrealistically large to fit
experimental data in some applications \cite{BK09}. It is shown in \cite{LE14}
that the entropy functional in \cite{BS11} does not directly yield classical
Boltzmann distributions $C_{i}(\mathbf{r})=C_{i}^{\text{B}}\exp\left(
-\beta_{i}\phi(\mathbf{r})\right)  $ as $v\rightarrow0$. It can be easily seen
from (9) that the entropy functional $F_{en}(\mathbf{C})$ in Eq. (8)
consistently yields Boltzmann distributions as $v_{i}\rightarrow0 $ for all
$i$. Our derivation of $F_{en}(\mathbf{C})$ does not employ any lattice models
but simply uses the volume equation (1). The functional $F_{en}(\mathbf{C})$
is a new modification of that in \cite{LE14}, where the classical Gibbs
entropy is now generalized to include all species (ions, water, and voids) in
electrolytes with the same entropy form. In fact, our $\Gamma(\mathbf{r)}$ is
an analytical extension of the void fraction $1-\Phi$ in Bikerman's excess
chemical potential \cite{BK09}, where all volume parameters $v_{i}$ (including
the bulk fraction $\Gamma^{\text{B}}$) are physical not empirical. The
adjustable parameter in our model is the correlation length $\lambda
\approx2a_{i}$ depending on the ionic species $i $ of interest \cite{LE14,L13}%
. The PF model was first proposed in \cite{BK09} without derivation and has
been shown to produce the results that are not only comparable to molecular
dynamics (MD) simulation or experimental data but also provide insight into
nonlinear properties of concentrated electrolytes and ionic liquids
\cite{SB12}. Here, we formally derive the PF model for general electrolytes
using a hard-sphere instead of lattice model with the steric potential
$S^{\text{trc}}(\mathbf{r})$ first introduced in \cite{L13}. As compared with
lattice models in \cite{BK09} and demonstrated, for example, in
\cite{LE14,GH13}, hard-sphere models significantly improve the agreement
between simulation and experiment.

(ii) The fourth-order PDE (7) is similar to those in \cite{S06,BS11} used in
previous papers \cite{LE14,L13}. However, the physical origin of the PDE is
different. In \cite{S06}, the global convolution is performed only on the
charge density of point-like counter ions in contrast to the potential
$\widetilde{\phi}(\mathbf{r})$ by Eq. (4) that is generated by all spherical
ions. In \cite{BS11}, a more general derivation for electrolytes or ionic
liquids with steric effects is given from a free energy function of a gradient
expansion of nonlocal electrostatic energy in terms of $\Delta\widetilde{\phi
}$. Eq. (7) corresponds to the first term in the expansion. Here, the
fourth-order PDE is derived directly from Maxwell's equation with the Yukawa
screening kernel. Our result does not depend on the convergence properties of
an expansion of nonlocal electrostatic energy.

(iii) Eq. (7) defines an dielectric operator $\widehat{\epsilon}=\epsilon
_{s}\epsilon_{0}\left(  1-\lambda^{2}\Delta\right)  $ that in turn implicitly
yields a dielectric function $\widetilde{\epsilon}(\mathbf{r})$ as an output
of solving Eq. (7) \cite{BS11,LE14}. A physical interpretation of the operator
was first introduced in \cite{BS11} to describe the nonlocal permittivity in a
correlated ionic liquid. The exact value of $\widetilde{\epsilon}(\mathbf{r})$
at any $\mathbf{r}\in$ $\Omega_{s}$ cannot be obtained from Eq. (7) but can be
approximated by the simple formula $\widetilde{\epsilon}(\mathbf{r}%
)\approx\epsilon_{0}+$ $C_{\text{H}_{2}\text{O}}(\mathbf{r}\mathbb{)(\epsilon
}_{s}-$ 1$)\epsilon_{0}/C_{\text{H}_{2}\text{O}}^{\text{B}}$ since the water
density function $C_{\text{H}_{2}\text{O}}(\mathbf{r}\mathbb{)}=C_{K+1}%
(\mathbf{r}\mathbb{)}$ is an output of Eq. (9). This formula is only for
visualizing (approximately) the profile of $\widehat{\epsilon}$ or
$\widetilde{\epsilon}$. It is not an input of calculation. The input is the
operator $\widehat{\epsilon}$ $=\epsilon_{s}\epsilon_{0}\left(  1-\lambda
^{2}\Delta\right)  $ (or the correlation length $\lambda$).

(iv) The factor $v_{i}/v_{0}$ multiplying $S^{\text{trc}}(\mathbf{r})$ in Eq.
(9) is a modification of the unity used in our previous work \cite{LE14}. The
steric energy $-\frac{v_{i}}{v_{0}}S^{\text{trc}}(\mathbf{r})k_{B}T$
\cite{LE14} of a type $i$ particle depends not only on the emptiness
($\Gamma(\mathbf{r)}=1-\sum_{i=1}^{K+1}v_{i}C_{i}(\mathbf{r})$) (or
equivalently crowding) at $\mathbf{r}$ but also on the volume $v_{i}$ of each
type of particle. If all $v_{i}$ are equal (and thus $v_{i}=v_{0}$), then all
particle species at any location $\mathbf{r}\in\Omega_{s}$ have the same
steric energy and the uniform particles are indistinguishable in steric
energy. The steric potential is a mean-field approximation of Lennard-Jones
(L-J) potentials that describe local variations of L-J distances (and thus
empty voids) between every pair of particles. L-J potentials are highly
oscillatory and extremely expensive and unstable to compute numerically.

(v) The global convolution in Eq. (5) may seem similar to those in
\cite{HB04,XJ12} but it is not. The Poisson equation (4) takes the place of
the Fourier-Lorentzian (FL) equation --- an integro-differential equation ---
in the previous work \cite{HB04,XJ12} in which the local potential
$\widetilde{\phi}(\mathbf{r}^{\prime})$ in Eq. (5) is replaced by a global
electric potential $\Phi(\mathbf{r}^{\prime})$. Moreover, the factor
$1/\lambda^{2} $ in Eq. (5) is replaced by $(\epsilon_{s}-\epsilon_{\infty
})/\lambda^{2} $, where $\epsilon_{\infty}$ and $\lambda$ are both adjustable
parameters. The choice of three parameters $\epsilon_{s}$, $\epsilon_{\infty}%
$, and $\lambda$ in the FL model is reduced to only one $\lambda$ here.

The nonlocal PF model is first compared with the modified PB model (mPB) of
Borukhov et al. \cite{BA97} in which ions are treated as cubes without
considering void and correlation effects. The classical PB model (with
$\lambda=v_{i}=0$ for all $i$, i.e., no size, void, and correlation effects)
produces unphysically high concentrations of anions near the charged wall at
$x=0$ as shown by the dashed curve in Fig. 1. The surface charge density is
$1e/(50$\r{A}$^{2})$ in contact with a 0.1 M C$_{4}$A aqueous electrolyte,
where the radius of both cations and anions is $a=4.65$ \r{A}\ in contrast to
an edge length of 7.5 \r{A}\ of cubical ions in \cite{BA97}, $e$ is the proton
charge, and $\epsilon_{s}=80$. The multivalent ions A$^{4-}$ represent large
polyanions adsorbed onto a charged Langmuir monolayer in experiments
\cite{BA97}. The dotted curve in Fig. 1 is similar to that of mPB in
\cite{BA97} and was obtained by the PF model with the size effect but without
voids and correlations, i.e., $\lambda=0$, $V_{K+2}=0$ (no voids), and
$v_{K+1}=v_{\text{H}_{2}\text{O}}=0$ (water is volumeless and hence
$\Gamma^{\text{B}}=1-\sum_{i=1}^{K}v_{i}C_{i}^{\text{B}}$ is the bulk water
volume fraction). The voids ($V_{K+2}\neq0$) and water molecules
($v_{\text{H}_{2}\text{O}}\neq0$) have slight effects on A$^{4-}$
concentration (because of saturation) and electric potential (because water
and voids have no charges) profiles as shown by the thin solid curves in Figs.
1 and 2, respectively, when compared with the dotted curves. However,
correlations (with $\lambda=2a$ \cite{BS11}) of ions have significant effects
on ion distributions as shown by the thick solid and dash-dotted curves in
Fig. 1, where the Stern layer is on the order of ionic radius $a$\ \cite{S24}
and the overscreening layer \cite{BS11} ($C_{\text{A}^{4-}}(x)\approx0$) of
excess coions ($C_{\text{C}^{+}}(x)$
$>$
$C_{\text{C}^{+}}^{\text{B}}=0.4$ M) is about 18 \r{A}\ in thickness. The
Stern layer is an output (not a prescribed condition) of our model. The
electric potentials $\phi(0)=$ 5.6 at $x=0$ and $\phi(11.5)=$ -1.97 $k_{B}T/e$
in Fig. 2 obtained by PF with voids and correlations deviate dramatically from
those by previous models for nearly 100\% at $x=0$ (in the Stern layer) and
70\% at $x=11.5$ \r{A}\ (in the screening layer) when compared with the
maximum potential $\phi(0)=$ 2.82 $k_{B}T/e$ of previous models.
\begin{figure}[t]
\centering\includegraphics[scale=0.7]{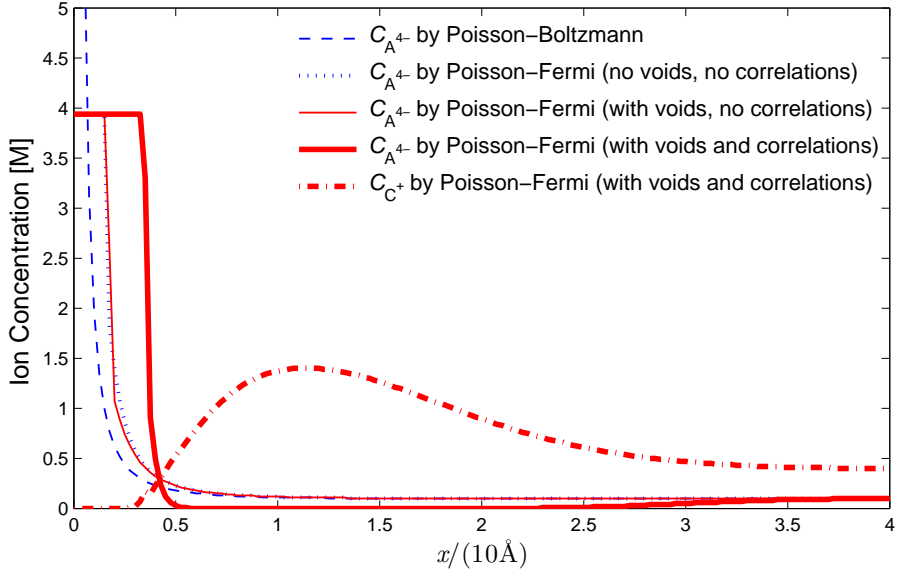}\caption{Concentration profiles
of anions $C_{\text{A}^{4-}}(x)$ and cations $C_{\text{C}^{+}}(x)$ obtained by
various models in a C$_{4}$A electrolyte solution with a positively charged
surface at $x=0$.}%
\end{figure}\begin{figure}[tt]
\centering\includegraphics[scale=0.7]{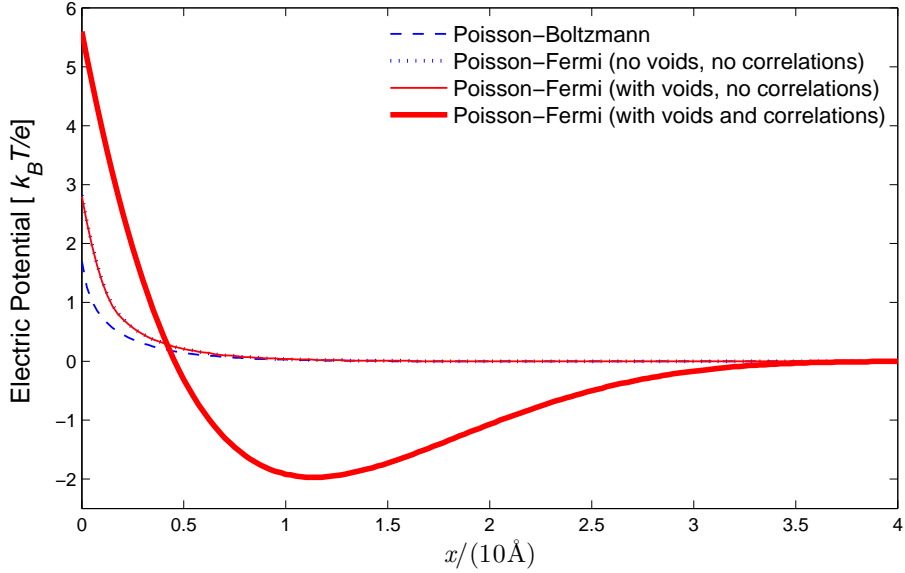}\caption{Electric potential
profiles $\phi(x)$.}%
\end{figure}  

We next show the size effect of different ions with voids using the crystal
structure of the potassium channel KcsA (PDB \cite{B02} ID 3F5W \cite{CJ10})
as shown in Fig. 3, where the spheres denote five specific cation binding
sites (S0 to S4) \cite{NB04}. The crystal structure with a total of $N=31268$
charged atoms is embedded in the protein domain $\Omega_{p}$ while the binding
sites are in the solvent domain $\Omega_{s}$. The exquisite selectivity of
K$^{+}$ (with $a_{\text{K}^{+}}=1.33$ \r{A}) over other cations such as
Na$^{+}$ ($a_{\text{Na}^{+}}=0.95$ \r{A}) by potassium channels is an
intriguing quest in channology. It can be quantified by the free energy ($G$)
difference of K$^{+}$ and Na$^{+}$ in the pore and in the bulk solution
\cite{NB04,NM88}. Experimental measurements \cite{NM88} showed that the
relative free energy \cite{NB04}
\begin{equation}
\Delta\Delta G(\text{K}^{+}\rightarrow\text{Na}^{+})=\left[  G_{\text{pore}%
}(\text{Na}^{+})-G_{\text{bulk}}(\text{Na}^{+})\right]  -\left[
G_{\text{pore}}(\text{K}^{+})-G_{\text{bulk}}(\text{K}^{+})\right]  \tag{11}%
\end{equation}
is greater than zero in the order of 5-6 kcal/mol unfavorable for Na$^{+}$.
The electric and steric potentials at the binding site S2 (as considered in
\cite{NB04}) can be calculated on the atomic scale using the following
algebraic formulas%
\begin{equation}
\phi_{\text{S2}}=\frac{1}{4\pi\epsilon_{0}}\left(  \frac{1}{6}\sum_{k=1}%
^{6}\sum_{j=1}^{N}\frac{q_{j}}{\epsilon_{p}(r)|c_{j}-A_{k}|}+\frac
{q_{\text{S2}}}{\epsilon_{b}a_{\text{S2}}}\right)  \text{, }S_{\text{S2}%
}^{\text{trc}}=\ln\frac{1-\frac{v_{\text{S2}}}{V_{\text{S2}}}}{\Gamma
^{\text{B}}},\tag{12}%
\end{equation}
where S2 = Na$^{+}$ or K$^{+}$ (the site is occupied by a Na$^{+}$ or a
K$^{+}$), $q_{j}$ is the charge on the atom $j$ in the protein given by
PDB2PQR \cite{DC07}, $\epsilon_{p}(r)=1+77r/(27.7+r)$ \cite{MM02}, $r=|c_{j}-$
$c_{\text{S2}}|$, $c_{j}$ is the center of atom $j$, $A_{k}$ is one of six
symmetric surface points on the spherical S2, $\epsilon_{b}=3.6$, and
$V_{\text{S2}}=1.5v_{\text{K}^{+}}$ is a volume containing the ion at S2. The
crucial parameter in (12) is the ionic radius $a_{\text{S2}}=0.95$ or $1.33$
\r{A}\ (also in $|c_{j}-A_{k}|$) that affects $\phi_{\text{S2}}$ very strongly
but $S_{\text{S2}}^{\text{trc}}$ weakly. We obtained $\Delta\Delta G=5.26$
kcal/mol in accord with the MD result in \cite{NB04}, where $G_{\text{pore}}%
($Na$^{+})=4.4$, $G_{\text{bulk}}($Na$^{+})=-0.26$ \cite{VB15},
$G_{\text{pore}}($K$^{+})=-0.87$, $G_{\text{bulk}}($K$^{+})=-0.27$ kcal/mol
\cite{VB15}, $G_{\text{pore}}($S2$)=q_{\text{S2}}\phi_{\text{S2}}%
-\frac{v_{\text{S2}}}{v_{0}}S_{\text{S2}}^{\text{trc}}k_{B}T$, $T=298.15$,
$\phi_{\text{Na}^{+}}=7.5$ $k_{B}T/e$, $\frac{v_{\text{Na}^{+}}}{v_{0}%
}S_{\text{Na}^{+}}^{\text{trc}}=0.23$, $\phi_{\text{K}^{+}}=-1.93$ $k_{B}T/e$,
$\frac{v_{\text{K}^{+}}}{v_{0}}S_{\text{K}^{+}}^{\text{trc}}=-0.59$, and
$C_{\text{Na}^{+}}^{\text{B}}=C_{\text{K}^{+}}^{\text{B}}=0.4$
M.\begin{figure}[t]
\centering\includegraphics[scale=0.7]{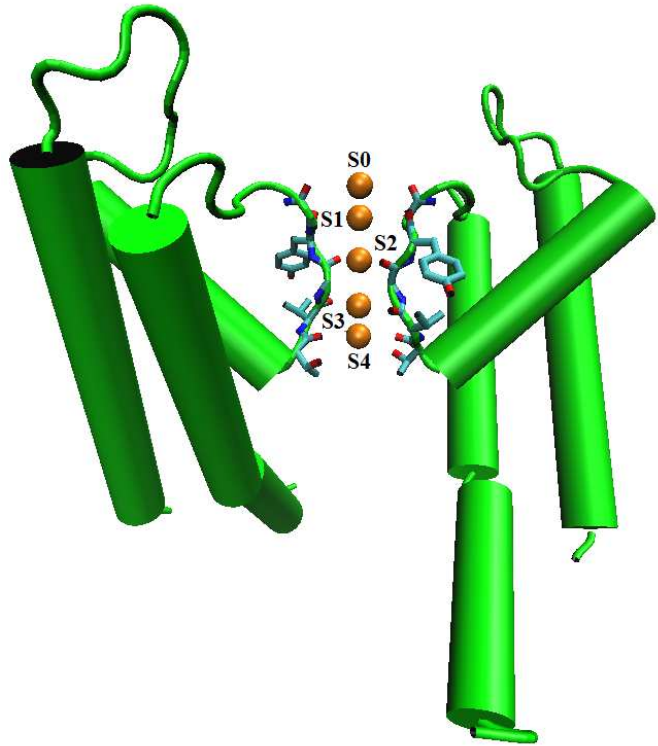}\caption{The crystal structure
of the K$^{+}$ channel KcsA (PDB ID 3F5W) \cite{CJ10} with five cation binding
sites S0, S1, S2, S3, and S4 \cite{NB04} marked by spheres.}%
\end{figure}

In summary, a nonlocal Poisson-Fermi model is proposed to describe global
electrostatic and steric effects that play a significant role of ionic
activities in electrolyte solutions especially in high field or large
concentration conditions. The model is based on Maxwell's field theory and
nonuniform hard spheres of all ions and water molecules with interstitial
voids. The Fermi-like distribution formula can describe the distribution of
nonuniform spherical ions and water molecules with interstitial voids. The
steric potential is a mean-field description of Lennard-Jones potentials
between particles. Poisson's equation is self-consistent with Fermi
distributions and global electrostatics. The present theory can be used to
describe complex functions of biological or chemical structures on both atomic
and macroscopic scales with far field bulk and boundary conditions.
Comparisons with experimental data are promising but incomplete.

This work was partially supported by the Ministry of Science and Technology,
Taiwan (MOST 103-2115-M-134-004-MY2 to J.-L. Liu) and the National Science
Foundation, USA (DMS-1226259 to D. Xie). We thank the Mathematical Biosciences
Institute at the Ohio State University, USA for support for our visits and get
together in Columbus in the fall of 2015.

\end{document}